\documentclass[apj]{emulateapj}

\usepackage{amssymb}

\usepackage{graphicx}
\usepackage{txfonts}
\usepackage{natbib}
\usepackage{lscape}

\newcommand{\msun}{{\rm\ M}_\odot}
\newcommand{\kms}{km~s$^{-1}$}

\newcommand\na{{New A}}%

\newcommand\actaa{{Acta Astron.}}%

\shorttitle{On the Progenitors of Galactic Novae}
\shortauthors{Darnley et al.}

\begin{document}

\title{On the Progenitors of Galactic Novae}

\author{M.~J. Darnley, V.~A.~R.~M. Ribeiro\altaffilmark{1,2}, M.~F. Bode, R.~A. Hounsell and R.~P. Williams\altaffilmark{3}}
\affil{Astrophysics Research Institute, Liverpool John Moores University, Egerton Wharf, Birkenhead, CH41~1LD, UK}
\email{M.Darnley@astro.ljmu.ac.uk}

\altaffiltext{1}{Astrophysics, Cosmology and Gravity Centre, Department of Astronomy, University of Cape Town, Private Bag X3, Rondebosch 7701, South Africa}
\altaffiltext{2}{South African Square Kilometre Array Fellow}
\altaffiltext{3}{Department of Physics, The University of Liverpool, Liverpool, L69~7ZE, UK}

\begin{abstract}
Of the approximately 400 known Galactic classical novae, only ten of them, the recurrent novae, have been seen to erupt more than once.  At least eight of these recurrents are known to harbor evolved secondary stars, rather than the main sequence secondaries typical in classical novae.  In this paper, we propose a new nova classification system, based solely on the evolutionary state of the secondary, and not (like the current schemes) based on the properties of the outbursts.  Using archival optical and near-infrared photometric observations of a sample of thirty eight quiescent Galactic novae we show that the evolutionary state of the secondary star in a quiescent system can predicted and several objects are identified for follow-up observations; CI~Aql, V2487~Oph, DI~Lac and EU~Sct.
\end{abstract}


\section{Introduction}\label{sec1}

A classical nova (CN) outburst occurs in an interacting binary system comprising a white dwarf (WD; the primary) and typically a late-type main-sequence star (the secondary) that fills its Roche Lobe \citep{1956ApJ...123...44C}.   Matter is transferred, usually via an accretion disk around the WD, from the secondary to the WD surface, becoming degenerate in the process. Once extensive CNO cycle nuclear burning commences under these degenerate conditions, a thermonuclear runaway occurs ejecting $\sim 10^{-5} - 10^{-4} \msun$ of matter from the system with velocities from hundreds to thousands of \kms\ \citep[see, e.g.][and references therein]{2008clno.book.....B}. These systems generally have orbital periods of a few hours \citep{2008clno.conf....1W} and recurrence times of a few $\times\;10^{3}-10^{6}$ years are often quoted \citep{1986ApJ...308..721T}.  CNe are traditionally sub-divided using solely their eruptive properties, typically speed class \citep{1957gano.book.....G}, or spectral class \citep{1992AJ....104..725W}.  More recently, novae have also been classified as ``disk'' or ``bulge'' novae, on the basis of their parent stellar population and location within the host galaxy \citep{1990LNP...369...34D,1992A&A...266..232D,1998ApJ...506..818D}.

Closely related to these systems are the recurrent novae (RNe), which, unlike CNe, are observed to recur on time-scales of order tens to a hundred years. However, it should be noted that the upper limit to this range is likely to be an observational effect based on the paucity of historical records. Such a short recurrence time-scale is usually attributed to a high mass WD, probably close to the Chandrasekhar limit, together with a high accretion rate \citep{1985ApJ...291..136S,2005ApJ...623..398Y}.  The orbital periods of RNe are less homogeneous than those of CNe, with periods ranging from a few hours to several hundreds of days \citep{2008ASPC..401...31A}.

Recurrent nova systems have been sub-divided into three general sub-classes \citep{2008ASPC..401...31A} via a combination of the properties of the eruption and via the properties of the progenitor system whilst at quiescence:

{\it RS~Oph/T~CrB systems} (hereafter the RS~Oph-class) are observed to contain red giant secondaries and hence have much longer orbital periods ($\sim$ a year), and typically smaller outburst amplitudes than CNe.  Their outbursts exhibit rapid declines from maximum, with large ejection velocities ($\ge 4000\;\mathrm{km\;s}^{-1}$).  These systems show evidence of an interaction between the ejecta and material from the pre-existing red giant wind.  The ejected mass from these systems is typically two orders of magnitude less than that observed from CN systems \citep[see papers within][]{2008ASPC..401.....E}.

{\it U~Sco systems} are observed to contain evolved main sequence or sub-giant secondaries, with orbital periods closer to those of CNe (hours up to of order a day).  They again exhibit rapid declines, being amongst the fastest declining novae observed, with very high ejection velocities (up to $10000\;\mathrm{km\;s}^{-1}$).  Their post-outburst spectra resemble those of the He/N sub-class of CNe \citep{1992AJ....104..725W}.  These systems eject a similar mass of material as RS Oph systems.

The {\it T~Pyx systems} (also CI~Aql and IM~Nor) are much more akin to CNe.  They exhibit short orbital periods and spectroscopically resemble Fe~II CNe \citep{1992AJ....104..725W}.  Their optical decline rate classifies them as moderately fast or slow novae.  The ejected mass in these systems is consistent with the range observed in CNe ($\mathrm{M}_{\mathrm{ej}}\sim 10^{-5}\msun$).  These systems are generally only distinguishable from CNe due to their shorter recurrence times, not by the properties of the progenitor system or the outburst.

The short inter-outburst time observed in RNe is likely due to some combination of a higher mass WD and an accretion rate greater than the typical for CN systems.  Indeed, both RS~Oph and U~Sco (amongst others from these classes) appear to have WDs close to the Chandrasekhar limit.  A number of authors \cite[see e.g.][]{2007ApJ...659L.153H,2011ApJ...727..124O,SumnerConf} have indicated that the WD mass may be increasing over time and these systems have been proposed as a Type Ia supernova (SN) progenitor candidate \citep[see e.g.][]{2008ASPC..401..150K}.  However, \citet{2011arXiv1107.4013M} reported that U~Sco could contain an O-Ne WD, rather than a C-O WD, and hence may not evolve in to a Type Ia SN as is widely predicted for this system.

Any CN that has been observed to erupt more than once is re-classified as a recurrent nova.  This approach has given us ten Galactic RNe amongst $\sim400$ known Galactic CNe from an underlying rate of $34^{+15}_{-12}$~year$^{-1}$ \citep{2006MNRAS.369..257D}.  Due to observational effects - which typically worsen the further back in time one looks - it is highly likely that a considerable proportion of these Galactic CNe are in fact (short time-scale) recurrents for which only a single outburst has been observed.  Either a previous outburst has been missed or the inter-outburst period is particularly long.  So it would be seem advantageous to better develop a nova classification system that relies more on the fundamental properties of the systems than on observational selection effects.

Over the past few years observations of a number of Galactic ``transients'' have shown properties consistent with being RNe, although they had never been seen in outburst previously.  For example, the nova V2487~Ophiuchi (Nova Oph 1998) was observed to have a rapid optical decline and plateau phase \citep{2002ASPC..261..629H,2002AIPC..637..381H}, suggestive of a RN.  Consultation of the Harvard College Observatory archival photographic collection revealed a previously unknown outburst on 1900 June 20 \citep{2009AJ....138.1230P}.  Additionally, the novae V1721~Aql \citep{2011arXiv1104.3068H}, V2491~Cyg \citep{2011arXiv1104.3482D}, KT~Eri \citep{RibKTTEri} and V2672~Oph \citep{2010MNRAS.tmp.1484M} and also the extragalactic nova M31N~2007-12b \citep{2009ApJ...705.1056B} have been shown to harbor secondary stars akin to either the U~Sco or RS~Oph systems, albeit with only one recorded outburst.

In this paper we will attempt to simplify the nomenclature somewhat by only classifying each system using the evolutionary state of the secondary; a main sequence star (MS-Nova), a sub-giant star (SG-Nova), or a red-giant branch star (RG-Nova).  We introduce such classifications in an attempt to avoid the accretion rate / WD mass / recurrence time degeneracy.  All the known U~Sco-class RNe would be placed into the SG-Nova group, all the known RS~Oph-class RNe into the RG-Nova group, the T~Pyx-class RNe and Classical Novae will populate the MS-Nova group.

By virtue of the geometry of these systems, the above classifications would effectively be the same as sub-dividing novae by orbital period, if one assumes that the secondaries fill their Roche lobes.  The MS-Novae would have orbital periods of order hours, the SG-Novae of order a day, and the RG-Novae of order a year.  Indeed \citet{1995CAS....28.....W} indicates that systems with orbital periods longer than eight hours should contain evolved secondaries.

In this paper we present a simple method which will allow differentiation of RG-Novae and SG-Novae from the MS-Novae population based solely on their quiescent (inter-outburst) broad-band optical and near-IR (NIR) properties.  The format of the remainder of this paper/letter is as follows: In Section~\ref{data} we briefly describe our method and present the data used to illustrate this technique; in Section~\ref{results} we present our results, in Section~\ref{discussion} we discuss our findings and present a number of predictions and finally, in Section~\ref{conc} we summarize our conclusions.

\section{Approach}\label{data}

The broad-band optical emission from all quiescent nova systems is a super-position of the emission from the three main components of the system; the WD, the accretion disk, and the secondary.  In all cases we would expect the WD's contribution in the optical to be negligible.  The contribution of the accretion disk to the emission is a combination of a number of factors including; accretion rate, disk size, system inclination and wavelength.  Whereas the contribution from the secondary is much more straightforward and simply depends upon the type (mass, age and metalicity) of the star and wavelength.

Both the SG-Novae and RG-Novae have much smaller optical outburst amplitudes than MS-Novae, although the energetics of the outbursts of all three systems are comparable.  In the RG-Nova systems this can be readily explained by the presence of a - highly luminous - red giant secondary, possibly with some additional contribution from a disk, especially in bluer optical filters.  For the SG-Nova systems the cause is not as clear-cut but must be some combination of a - luminous - sub-giant secondary and an accretion disk, possibly experience a high mass accretion rate.

The broad-band NIR emission from quiescent RG-Nova and SG-Nova systems should be dominated by the emission from the cool secondary stars, with a small contribution from the accretion disk.  At these wavelengths, the accretion disk emission may still be relatively significant for MS-Nova systems due to the low luminosity of the dwarf secondary star.

As such, we propose that the position of a quiescent nova system on a color-magnitude diagram can be broadly predicted.  Each system would be expected to lie at such a position that it appears more luminous and bluer (hotter) than the secondary in the system, with the magnitude of the luminosity and temperature increase being a function of both the accretion rate, system inclination and - of course - wavelength.  Conversely, by plotting quiescent nova systems on a color-magnitude diagram one can identify the type of secondary and hence predict whether a given system may be an RG-Nova, SG-Nova or MS-Nova system.

In Table~\ref{tb:one} we present quiescent broad-band optical and NIR photometric data for a sample of Galactic novae (and a number of extragalactic examples).  These data have been collated from a wide range of sources, with all the NIR data taken from the Two Micron All Sky Survey \citep[2MASS;][]{2006AJ....131.1163S} either directly from the catalog or indirectly through additional sources.  In Table~\ref{tb:two} we provide a summary of the best determination of the line of sight extinction and distance to each system as well as the system inclination (if known).  This catalog of Galactic novae was compiled by virtue of the existence of quiescent photometry in at least two optical bands, and reasonable determinations of both distance and extinction.

\begin{table*}
\caption{Distance and extinction towards quiescent nova systems and inclination and orbital periods (if known) of those systems.  Here, all MMRD determinations are made using that calibrated by \protect{\citet{2000AJ....120.2007D}}.\label{tb:two}}
\begin{center}
\begin{tabular}{llllll}
\hline
\hline
Nova          & Distance               & Distance      & Extinction           & Inclination     & Orbital \\
Name          & (kpc)                  & Determination & $E_{B-V}$ (mag)        & (degrees)       & Period (h, 1) \\
\hline
CI Aql        	& 5.0$^{+5.0}_{-2.5}$ (2) & MMRD          	& 0.85$\pm$0.3 (2)     	& $68\pm1$ (3)					& 14.84\\
V356 Aql      	& 1.7 (4)              & Gal. rotation         	& 0.70 (5)             		& - & - \\
V528 Aql      	& 2.4$\pm$0.6 (5)      & Various 	& 0.84$\pm$0.19 (5)    	& - & -\\
V603 Aql      	& 0.33 (5)             & Ex. parallax          	& 0.08 (6)             		& 16 (7) 				& 3.324\\
V1229 Aql     	& 1.73 (5)             & Various          	& 0.52$\pm$0.13 (5)    	& - & -\\
V1721 Aql     	& $2.2\pm0.6$ (8)        		& MMRD          	& $3.7\pm0.1$ (8)      	& - & -\\
T Aur         	& 0.96$\pm$0.22 (9)      		& Ex. parallax  	& 0.39$\pm$0.08 (5,6)  	& 57/68 (10,11)			& 4.906\\
IV Cep        	& 2.05$\pm$0.15 (5)      		& Various         	& 0.53$\pm$0.03 (5)    	& - & -\\
V394 CrA      	& 10$^{+15}_{-3}$ (2)       		& MMRD          	& 0.20$\pm$0.20 (2)    	& -					& 1.52 d (2) \\
T CrB         	& 0.9$\pm$0.2 (2)        		& MMRD          	& 0.10$\pm$0.10 (2)    	& -					& 227.57 d\\
V407 Cyg      	& 2.7$\pm$0.3 (12)        		& Mira P-L      	& 0.57$\pm$0.02 (12)    	& - 					& 43$\pm$5 yr (13) \\
V476 Cyg      	& 1.62$\pm$0.12 (9)      		& Ex. parallax  	& 0.27$\pm$0.13 (5)    	& 65 (7) & -\\
V1500 Cyg     	& 1.5$\pm$0.2 (9)        		& Ex. parallax  	& 0.43$\pm$0.08 (5,6)  	& 55 (7) 				& 3.351\\
V1974 Cyg     	& 1.77$\pm$0.11 (14)     		& Various          	& 0.32$\pm$0.01 (15)   	& 38.7$\pm$2.1 (14) 	& 1.950\\
V2491 Cyg     	& 13.3$\pm$0.8 (16,17)                	& MMRD          	& 0.3$\pm$0.1 (16,17)            		& $80^{+3}_{-12}$ (17) & -\\
HR Del        	& 0.76$\pm$0.13 (9)     	 	& Ex. parallax  	& 0.15 (5,6)           		& 38 (18) 				& 5.140\\
KT Eri        	& 6.5 (19)               		& MMRD          	& 0.08 (19)             		& $53^{+6}_{-7}$ (20) 	& 737 d (21) \\
DN Gem        	& 0.45$\pm$0.07 (5)      		& MMRD          	& 0.08$\pm$0.03 (5,6)  	& 30 (7) 				& 3.068\\
DQ Her        	& 0.40$\pm$0.06 (9)      		& Ex. parallax  	& 0.08 (6)             		& 81$\pm$4 (9) 		& 4.647 \\
V533 Her      	& 0.56$\pm$0.07 (9)      		& Ex. parallax  	& 0.03 (6)             		& 62 (7) 				& 3.53 \\
CP Lac        	& 1.0$\pm$0.1 (5)        		& Ex. parallax         	& 0.48$\pm$0.03 (5)    	& -					& 0.15 d (22) \\
DK Lac        	& 3.9$\pm$0.5 (9)        		& Ex. parallax         	& 0.39$\pm$0.06 (5)    	& - & -\\
DI Lac        	& $2.25\pm0.25$ (23)     		& Model         	& 0.41 (23)            		& $<18$ (23) 			& 13.050\\
IM Nor        	& 3.4$^{+3.4}_{-1.7}$ (2)    	        & MMRD          	& 0.80$\pm$0.20 (2)    	& - 					& 2.46\\
RS Oph        	& 1.4$^{+0.6}_{-0.2}$ (24,25)	        & Various         	& 0.70$\pm$0.10 (26)   	& 39$^{+1}_{-10}$ (27) 	& 455.72 d\\
V849 Oph      	& 3.1 (5)                		& Various          	& 0.27 (28)            		& -					& 4.146\\
V2487 Oph     	& 12$^{+13}_{-2}$ (2)       		& MMRD          	& 0.50$\pm$0.20 (2)    	& -					& $\sim$1 d (2) \\
GK Per        	& 0.46$\pm$0.03 (9)      		& Ex. parallax  	& 0.29 (6)             		& 70 (29) 				& 47.923 \\
RR Pic        	& 0.4 (5)                		& Ex. parallax         	& 0.02 (6)             		& 70 (7) 				& 3.481\\
CP Pup        	& 1.5 (5)                		& Ex. parallax         	& 0.25$\pm$0.06 (5,6)  	& - 					& 1.474\\
T Pyx         	& 4.5$\pm$1.0 (30)        		& Various          	& 0.5$\pm$0.1 (30)    	& -					& 1.829\\
U Sco         	& 12$\pm$2 (2)           		& MMRD          	& 0.20$\pm$0.10 (2)    	& 80 (31) 				& 29.53\\
V745 Sco      	& 7.8$\pm$1.8 (2)        		& MMRD          	& 1.00$\pm$0.2 (2)     	& -					& 510$\pm$20 d (2) \\
EU Sct        	& 5.1$\pm$1.7 (5)        		& Various         	& 0.84$\pm$0.19 (28)    	& - & - \\
FH Ser        	& 0.92$\pm$0.13 (9)      		& Ex. parallax  	& 0.74 (5,6)           		& 58$\pm$14 (9) & -\\
V3890 Sgr     	& 7.0$\pm$1.6 (2)        		& MMRD          	& 0.90$\pm$0.30 (2)    	& -					& 519.7$\pm$0.3 d (2) \\
LV Vul        	& 0.92$\pm$0.08 (9)      		& Ex. parallax  	& 0.55 (28)            		& - & -\\
NQ Vul        	& 1.6$\pm$0.8 (9)        		& Ex. parallax  	& 0.81 (28)            		& - & -\\
\hline
LMCN 2009-02a & $48.1\pm2.3_{r}\pm2.9_{s}$ (32) & Various     & $0.19\pm0.06$ (33)    & - & -\\
M31N 2007-12b & $769^{+36}_{-35}$ (34)    & Cepheid       & $0.35\pm0.25|_{\mathrm{sys}}$ (35) & - & - \\
\hline
\end{tabular}
\end{center}
1. Unless otherwise stated, all periods are in hours and are from \citet{2008clno.conf....1W}, 2.~\citet{2010ApJS..187..275S}, 3.~Sahman et al. (in prep), 4.~\citet{1941ApJ....93..417M}, 5.~\citet{1981PASP...93..165D}, 6.~\citet{1994A&AS..104...79B}, 7.~\citet{1987MNRAS.227...23W}, 8.~\citet{2011arXiv1104.3068H}, 9.~\citet{1995MNRAS.276..353S}, 10.~\citet{1984A&AS...57..385R}, 11.~\citet{1980MNRAS.192..127B}, 12.~\citet{1990MNRAS.242..653M}, 13.~\citet{1990MNRAS.242..653M}, 14.~\citet{1997A&A...318..908C}, 15.~\citet{1993A&A...277..103C}, 16.~\citet{2010arXiv1009.0822M}, 17.~\citet{2011MNRAS.412.1701R}, 18.~\citet{1983ApJ...273..647S}, 19.~\citet{2009ATel.2327....1R}, 20.~\citet{2011PhDT.........1R}, 21.~\citet{RibKTTEri}, 22.~\citet{2006PASP..118..687P}, 23.~\citet{2003AJ....125..288M}, 24.~\citet{1987rorn.conf..241B}, 25.~\citet{2008ASPC..401...52B}, 26.~\citet{1987rorn.conf...51S}, 27.~\citet{2009ApJ...703.1955R}, 28.~\citet{1994AJ....108..639S}, 29.~\citet{1986MNRAS.222...11W}, 30.~\citet{2011arXiv1108.3505S}, 31.~\citet{2000ApJ...534L.189H}, 32.~\citet{2006ApJ...652.1133M}, 33.~Bode et al. (in prep), 34.~\citet{1990ApJ...365..186F}, 35.~\citet{2009ApJ...705.1056B}.
\end{table*}

\section{Results}\label{results}

\begin{figure*}
\begin{center}
\includegraphics[width=0.475\textwidth]{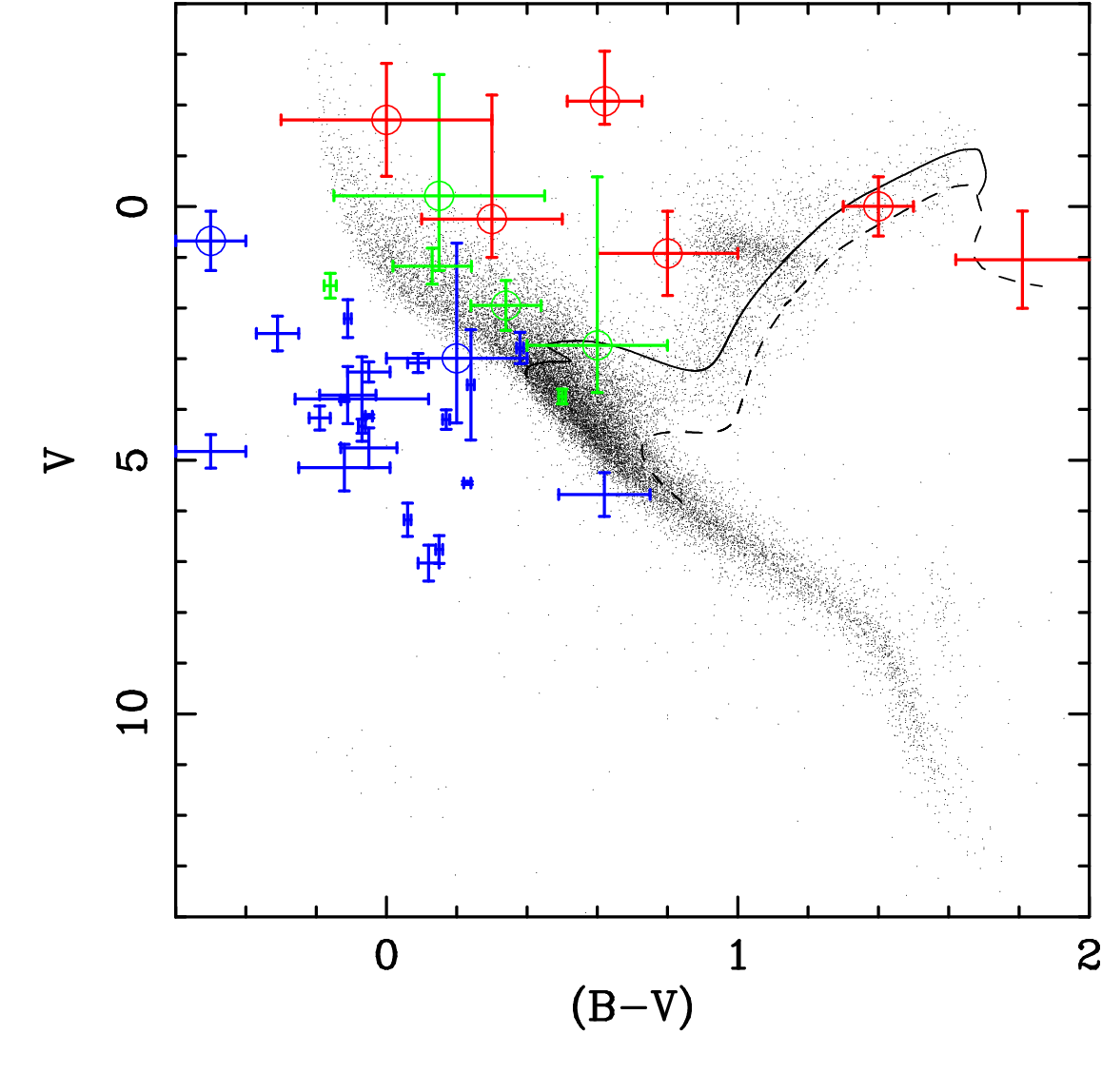}
\includegraphics[width=0.475\textwidth]{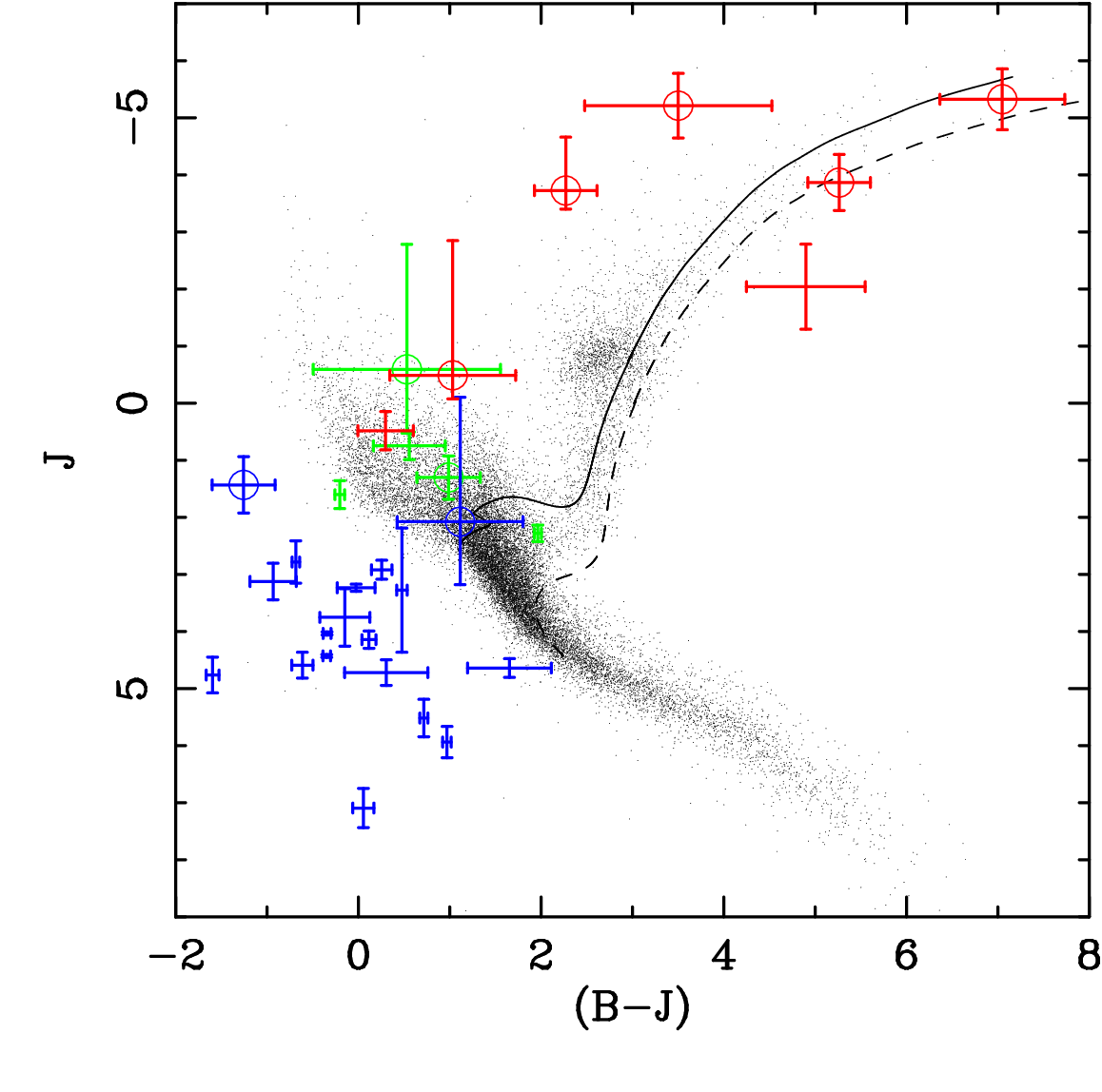}\\
\includegraphics[width=0.475\textwidth]{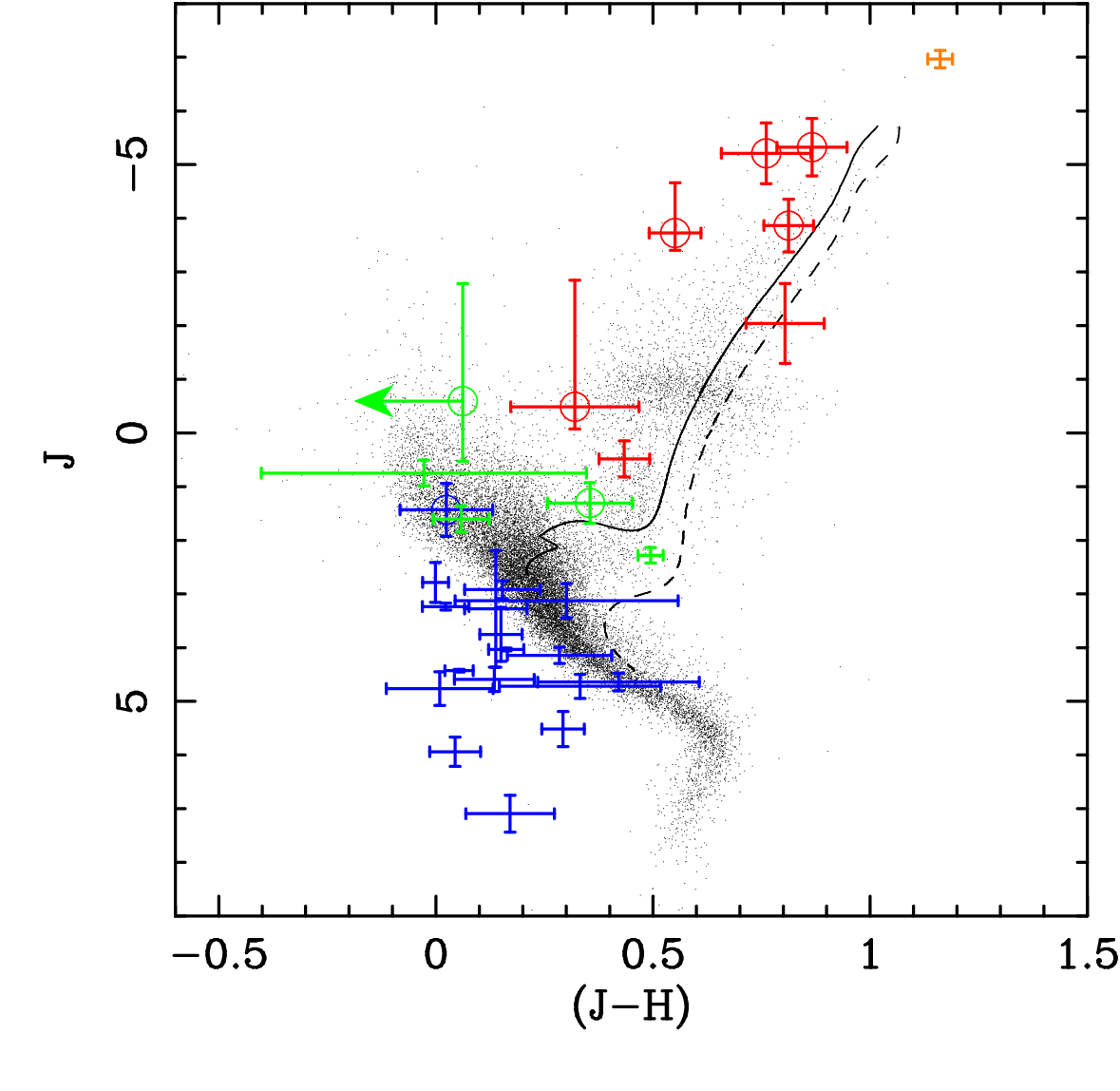}
\includegraphics[width=0.475\textwidth]{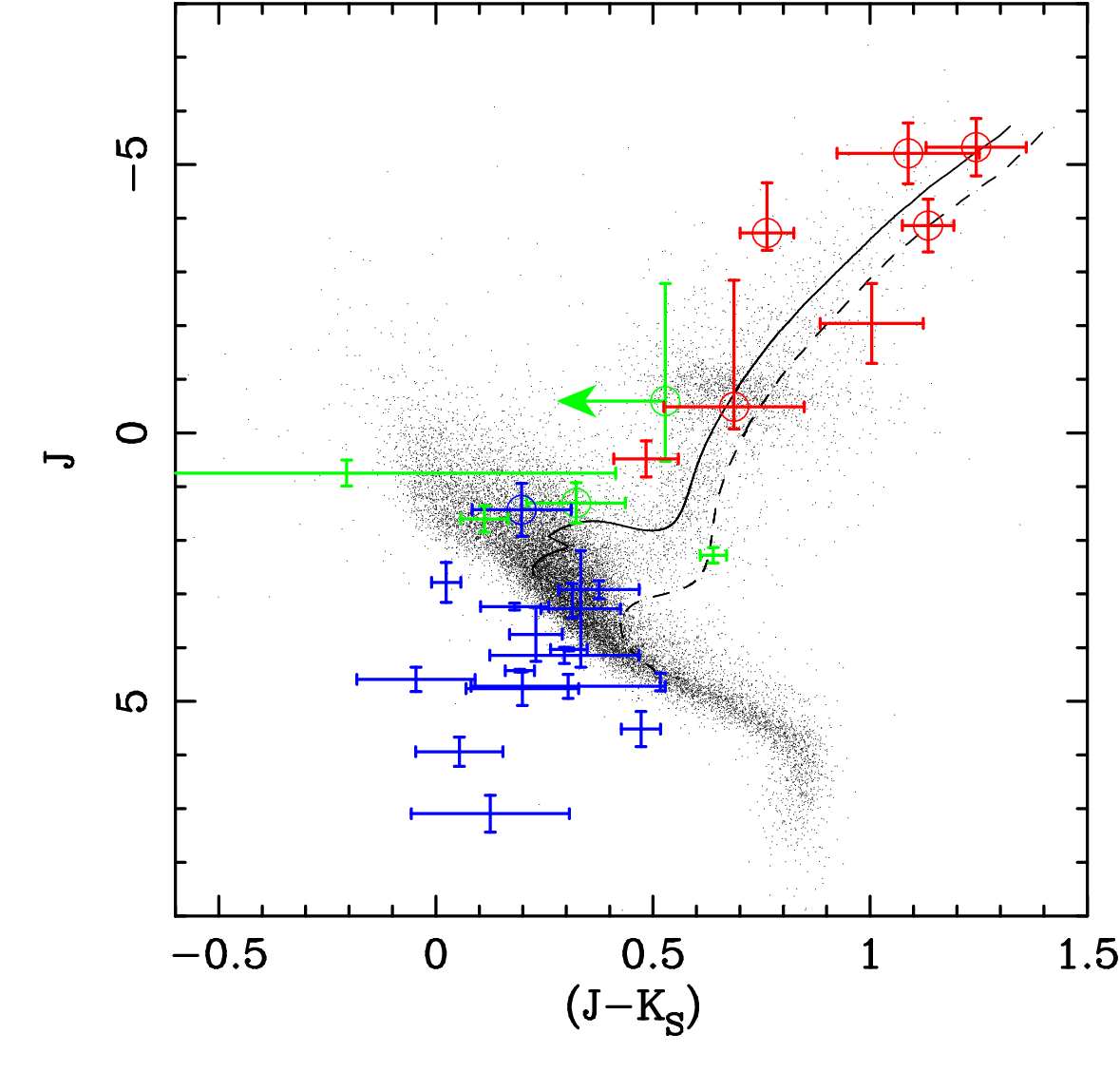}
\end{center}
\caption{Color-magnitude diagrams showing stars from the {\it Hipparcos} data set \citep{1997ESASP1200.....P}, with parallax and photometric errors $<10\%$. $BVI$ photometry is taken directly from the {\it Hipparcos} catalog, $R$ photometry is taken from the {\it NOMAD1} data set and NIR photometry from the 2MASS catalog.  The positions of the novae shown in Table \ref{tb:one} are plotted.  Dark blue points represent MS-Novae, green and red points show members or candidates of the SG-Novae class and RG-Novae class respectively, and the orange point shows the symbiotic nova V407~Cyg.  The known recurrent novae in this sample have been identified by an additional circle.  The black dashed line shows the evolutionary track of a 1M$_{\odot}$ Solar-like star, the solid line a 1.4M$_{\odot}$ Solar-like star \protect{\citep{2004ApJ...612..168P}}.}
\label{fig:one}
\end{figure*}

In Figure~\ref{fig:one} we present a range of color-magnitude plots.  The positions of all the novae listed in Table~\ref{tb:one} are plotted with appropriate error bars (or error estimations).  For comparison, we have also plotted the local Galactic stellar population, generated from the {\it Hipparcos} catalog \citep{1997ESASP1200.....P}.  Only those stars that have both photometric and parallax errors $<10\%$ have been selected from the {\it Hipparcos} data.  The $BVI$ photometry for these {\it Hipparcos} stars has been taken directly from that catalog, while the $R$ and NIR photometry has been generated by cross-correlating the {\it Hipparcos} catalog with the  {\it NOMAD1} data set \citep{2004AAS...205.4815Z} and the 2MASS cataogue respectively.  Included in each figure are the stellar evolutionary tracks of a 1M$_{\odot}$ Solar-like star and a 1.4M$_{\odot}$ Solar-like star \citep{2004ApJ...612..168P} from the zero-age main-sequence up to the tip of the red giant branch.  The majority of the nova secondary masses are expected to be $<1\mathrm{M}_{\odot}$, and for stable mass transfer the mass of the secondary must be $<\mathrm{M}_{\mathrm{WD}}$.  We have provided four alternative color-magnitudes diagrams: a standard diagram $V$ vs $B-V$; as well as $J$ vs $B-J$ to particulary focus on the luminosity difference between the disk and secondary; and a pair of NIR diagrams to focus specifically on the secondary.

Taking each plot from Figure~\ref{fig:one} in turn:

\begin{itemize}
\item $V$ vs $B-V$: This plot was included as this is the ``standard'' color magnitude diagram and the most commonly used filters, particularly for observations of novae.  Although there does seem to be some separation by position of the three secondary sub-types of nova systems, it is not overly clear.  Unless the secondary star is particularly bright, or the system is very close to edge on, the $B$ and $V$ photometry is strongly influenced by the accretion disk, and as such gives little information regarding the nature of the secondary star.  As can be seen, a number of RG-Novae (red points) are inter-dispersed amongst the SG-Novae (green points) probably due to a combination of low system inclination angles and high accretion rates.  However, the MS-Novae do appear to be separated from the SG-/RG-Novae by the main sequence itself, with the MS-Novae all lying on the blue-side.
\item $J$ vs $B-J$: This plot was included as it specifically samples both the secondary star ($J$) and the accretion disk ($B$) giving some indication of the relative luminosity of the accretion disk in each system, and hence the accretion rate.  As in the $V$ vs $B-V$ plot, a number of points have been redistributed significantly ``blue-wards'' from the position of the secondary.
\item $J$ vs $J-H$: The first of two NIR only color-magnitude diagrams.  Such plots are expected to focus strongly on the secondary star which is expected to dominate the emission at these wavelengths for all but the lowest luminosity secondaries.  As can be seen from the plot in Figure~\ref{fig:one} there is a clear separation between the RG-Novae and the MS-Novae, and possibly a small distinct region belonging to the SG-Novae.
\item $J$ vs $J-K_{S}$: As expected, this plot is very similar to the $J-H$ plot.  This plot is slightly let down by the lower sensitivity of the 2MASS $K_{S}$ data-set; however, the separation between the three sub-classes is just as well defined.
\end{itemize}

\begin{figure*}
\begin{center}
\includegraphics[width=0.475\textwidth]{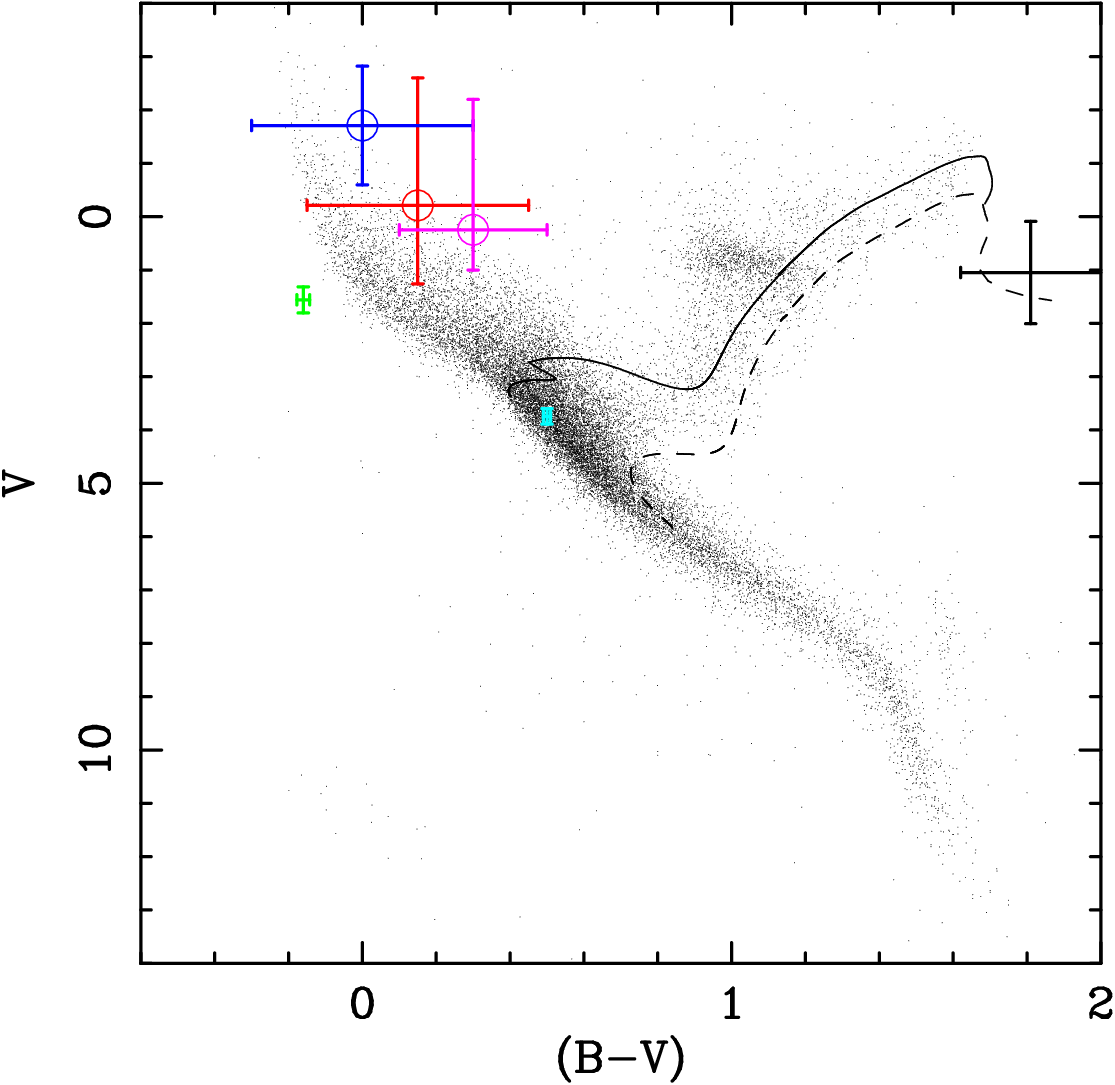}
\includegraphics[width=0.475\textwidth]{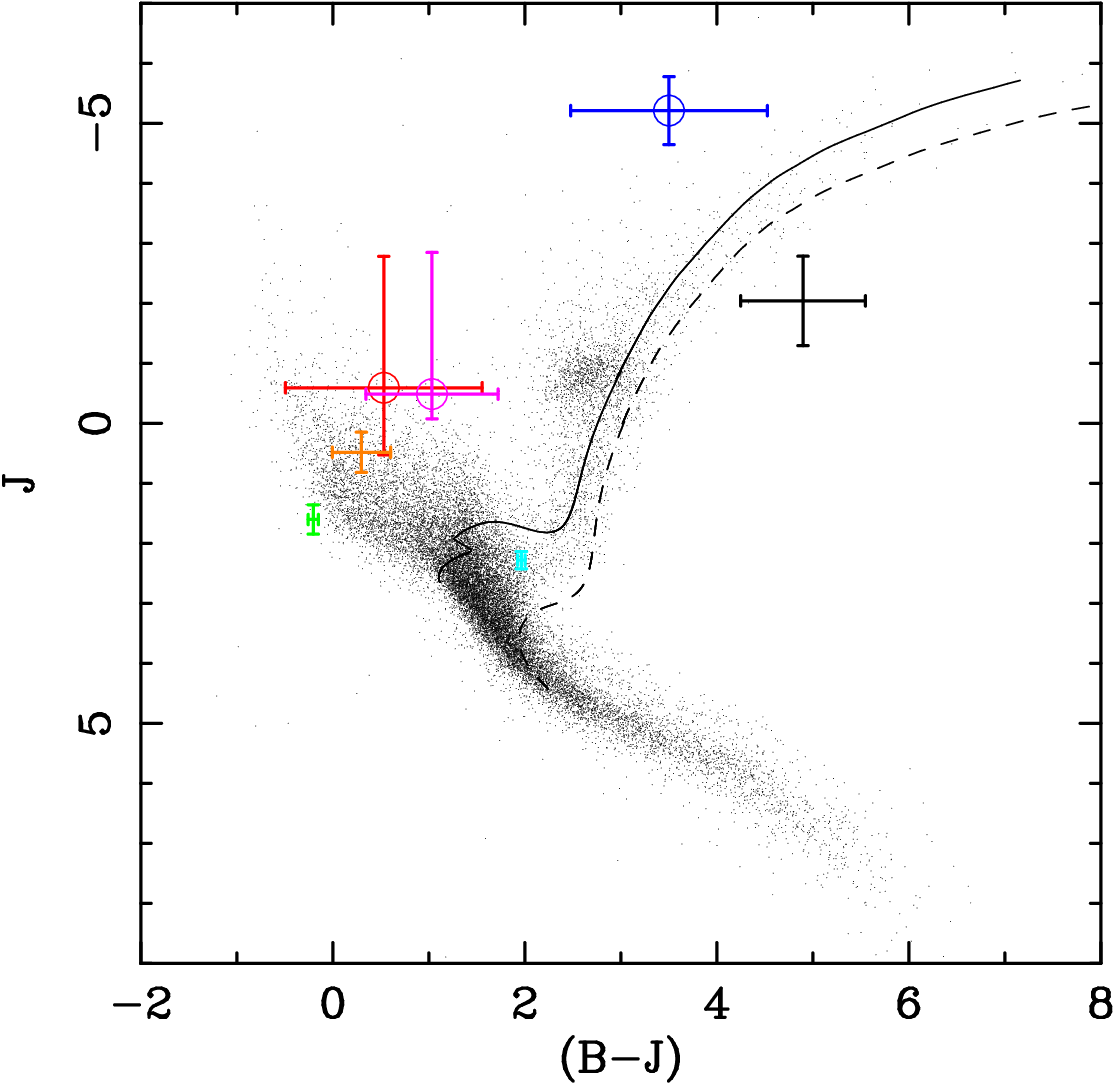}\\
\includegraphics[width=0.475\textwidth]{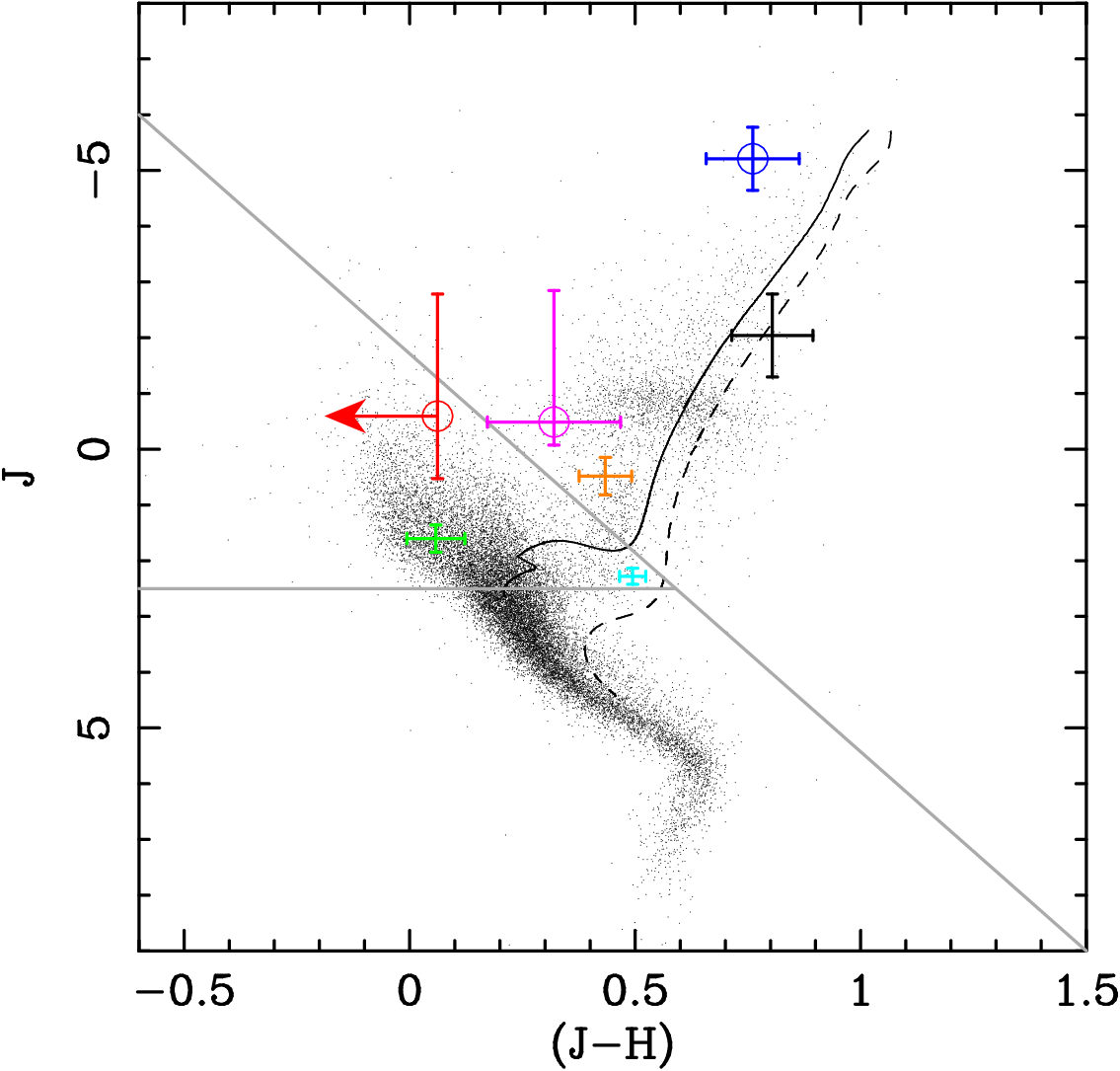}
\includegraphics[width=0.475\textwidth]{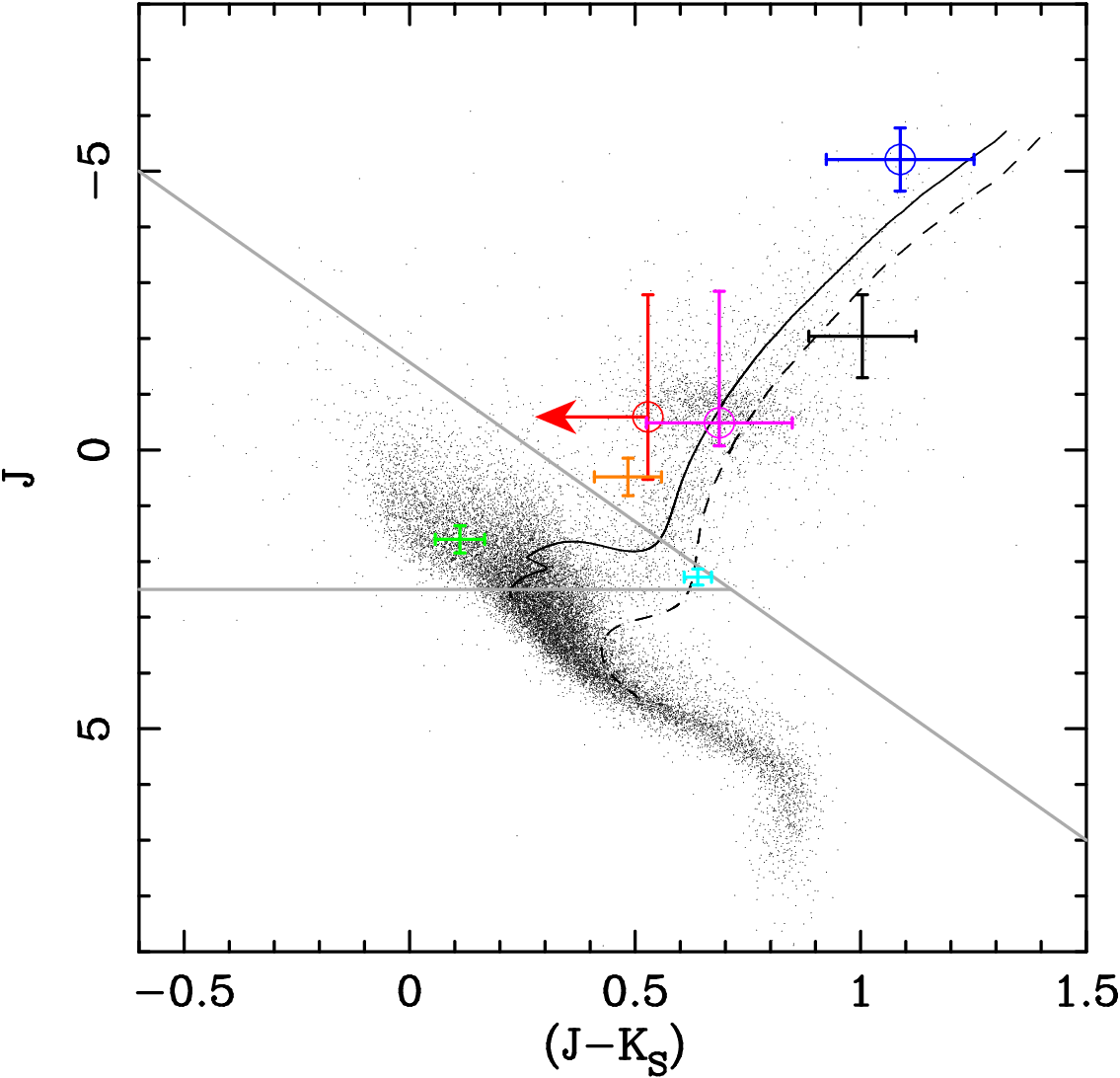}
\end{center}
\caption{As Figure~\ref{fig:one} but only identifying the specific novae discussed in the text.  The following colors are used to identify specific novae: CI~Aql - red; KT~Eri - orange; DI~Lac - green; V2487~Oph - magenta; GK~Per - cyan; EU~Sct - black; and V3890~Sgr - blue.  The solid gray lines in the NIR plots indicate approximate boundaries between the MS-Novae, SG-Novae and RG-Novae.}
\label{fig:two}
\end{figure*}

The plots created for Figure~\ref{fig:one} are also reproduced in Figure~\ref{fig:two}, however, in this figure we only plot (and identify) a selection of specific novae.  We will now briefly discuss each of these selected novae.

\subsection{CI~Aquil\ae}

The recurrent nova CI~Aql is occasionally placed within the T~Pyx-class of RNe due to both its recurrent nature and the properties of the outbursts \citep{2008ASPC..401...31A}, and as such, we would expect to observe this system within the cluster of MS-Novae.  However, on closer inspection, the quiescent photometry of the system itself is more akin to those of the SG-Novae (see Figure~\ref{fig:one} and the red points within Figure~\ref{fig:two}).  Whilst this system is undetected in the 2MASS $H$ and $K_{S}$ data, the upper limits from these filters coupled with the $J$ photometry indicate a luminous and blue system.  Couple this with the relatively long orbital period \citep[14.84 hours, see Table 2.5 in][]{2008clno.conf....1W} and with the short recurrence time of the outburst, which indicate that this system is a member of the U~Sco-class of RNe progenitor systems, i.e. a SG-Nova.

\subsection{KT Eridani}

As previously reported in \citet{RibKTTEri}, it has been shown that the central system of KT~Eri is likely to harbor a red giant secondary, less evolved than those in the RS~Oph and T~CrB systems.  Whilst \citet{RibKTTEri} only reported the analysis of the NIR quiescent data, here we also discuss the optical data.  The position of KT~Eri on a color-magnitude diagram can be seen in Figures~\ref{fig:one} and \ref{fig:two} (the orange points in the former).  It can be seen that this system has a strong blue excess, most likely due to a high accretion rate.  This implies an elevated accretion rate for the system and hence a relatively short inter-outburst time.  The similarities between the outbursts of KT~Eri and the known recurrent nova LMCN 2009-02a are discussed by both Bode et~al. (in prep) and \citet{RibKTTEri}.

\subsection{DI~Lacert\ae}

DI~Lac is previously classified as a CN, following a single recorded outburst in 1910 - hence most likely a MS-Nova.  Whilst the available ($BVJHK_{S}$) quiescent photometry doesn't exclude this conclusion, it is interesting to note that DI~Lac, is the brightest (by $\sim2$ magnitudes) of the MS-Novae (see the dark blue points in Figure~\ref{fig:one} and the yellow in Figure~\ref{fig:two}).  This system is around the same luminosity, in all filters, as  the U~Sco system, although the former is always slightly bluer.  As such, we cannot rule out the possibility that this system may be a SG-Nova, possibly with a less evolved secondary than U~Sco or with a higher accretion rate.  The quoted orbital period for DI~Lac is 13.05 hours \citep{2008clno.conf....1W} significantly longer than that of a typical MS-Nova, and similar to the ~day-long periods of SG-Nova systems.

\subsection{V2487~Ophiuchus}

Following analysis of the light curve of the 1998 outburst of V2487 Oph, \citet{2002ASPC..261..629H} identified this system as a U~Sco type (SG-Nova) nova, as a possible recurrent and Type Ia SN progenitor (although the simulations included a MS secondary).  \citet{2009AJ....138.1230P} then confirmed the recurrent nature with the discovery of a previous outburst in 1900.  The optical photometry of V2487~Oph is broadly supportive of the conclusions from \citet{2002ASPC..261..629H} (see Figure~\ref{fig:one} and magenta points in Figure~\ref{fig:two}).  However, when one looks at this system in the NIR (see again Figures~\ref{fig:one} and \ref{fig:two}), the system is clearly both luminous and red.  In fact the NIR is consistent with a system containing a secondary that is climbing the RGB, a horizontal branch star also cannot be ruled out (from the photometry alone).  The $(B-J)$ color of the system indicates that there is a strong blue component in the system in addition to the secondary star.  As such, the conclusion we draw is that V2487~Oph is a RG-Nova, whose secondary is less luminous/evolved than that of RS~Oph/T~CrB.  V2487~Oph is a system akin to KT~Eri, containing a very luminous accretion disk and hence has the high accretion rate required for a short inter-outburst timescale.

\subsection{GK~Persei}

We also briefly discuss the nova GK~Per, not because we disagree with previous determinations of the nature of this system, but because it is an interesting object in its own right.  The secondary in the GK~Per system is a K2IV star, making this system a (unusually, non-recurrent) SG-Nova.  Additionally the strength of the magnetic field of the WD places GK~Per into the intermediate polar \citep[or DQ~Her;][]{2008clno.conf....1W} class of objects.  As such, any accretion disk within the system is possibly less extensive than that within a typical CN and, as the system is close to edge-on, the effect of any disk on the quiescent photometry is diminished further.  Given this, and the high luminosity of the sub-giant secondary, we would expect the quiescent photometry to clearly indicate the H-R diagram position of the secondary star. As can be seen by the cyan points within Figure~\ref{fig:two} the GK~Per system can be found on the sub-giant branch in all four plots.

\subsection{V3890~Sagittarii}

Long known as a RS~Oph-class RN, V3890~Sgr lies in an interesting position on a standard $V$ versus $B-V$ color magnitude diagram (see the top-left plots within Figure~\ref{fig:one}, the blue point in Figure~\ref{fig:two}), that is grouped with the SG-Nova systems.  However, on inspection of the NIR plots (bottom row of Figure~\ref{fig:one} and \ref{fig:two}) it is clear that this system is more akin to the RG-Nova systems, as expected.  Such an apparent discrepancy can be easily resolved if V3890~Sgr has a particularly high accretion rate and hence a highly luminous disk.  The contribution of the disk to the total flux could also be boosted if the system orientation is close to face on.

\subsection{EU~Scutum}

Nova EU~Sct also lies in an interesting position on both the $V$ vs. $B-V$ color magnitude plot (see the top-left plot within Figure~\ref{fig:one}, green points in Figure~\ref{fig:two}) - that is significantly redwards of the RGB.  However, on inspection of the NIR diagrams, the position of this quiescent nova is consistent with the system containing a red giant secondary ($M_{J}=-2.0\pm0.7$, $M_{H}=-2.8\pm0.7$, $M_{K_{S}}=-3.0\pm0.7$).  Whilst it is possible that this contradiction could be explained by suspect $B$-band photometry, it is also possible that the extinction towards this system ($E_{B-V}=0.8\pm0.2$) has been severely underestimated.  Hence, given its quiescent photometry and small outburst amplitude, we conclude that EU~Sct is likely to be an RG-Nova.  This conclusion is in line with those of \citet{1994MNRAS.266..761W} and \citet{2003MNRAS.340.1011W}.

\section{Discussion}\label{discussion}

As can be seen in Figure \ref{fig:one}, when quiescent novae are plotted on color-magnitude diagrams, the position of a given system is a strong function of both the evolutionary state of the secondary star and the luminosity of any accretion disk.  When we focus only on NIR observations, any dependency on the accretion disk is minimized, and in many cases is negligible.  Those systems known to harbor red-giant secondary stars are correlated strongly with the red-giant branch.  Those containing main-sequence secondaries are clustered blue-wards of the main sequence (due to the effect of the accretion disk) yet are less luminous than the (current) position of the local Galactic sub-giant branch.  Whilst less well defined, the locus of the sub-giant secondary systems is typically more luminous than the sub-giant branch and significantly bluer than the red-giant branch.

This work has allowed us to reach a number of conclusions and predictions.  Firstly, we can identify a number of systems structurally akin to either U~Sco (SG-Novae) or RS~Oph (RG-Novae) that are only known to have undergone a single outburst.  These SG-Novae are V1721~Aql and V2491~Cyg; the RG-Novae are KT~Eri, EU~Sct and M31N 2007-12b.  However, the question of whether these are recurrent novae with missed outbursts or part of a population of long inter-outburst timescale evolved systems is one we cannot address here as a short recurrence time also requires a high mass WD.  However, we would strongly urge searches of observational archives to uncover any missed outbursts (as was done for V2487~Oph).  We predict that the secondary of the recurrent nova CI~Aql is a sub-giant star, akin to the secondary in the U~Sco system, and that the secondary of the recurrent V2487~Oph is a red-giant similar to that in the KT~Eri system.  There is also some evidence that the secondary in DI~Lac is also a sub-giant star.

Whilst preparing the catalog of novae analyzed in this paper, it quickly became clear that reliable, multi-color, quiescent photometry is lacking for a large number of Galactic novae.  Additionally, reliance on a single epoch of quiescent photometry is problematic due to the large range in quiescent luminosity exhibited by many systems.

Whilst this technique has been shown to produce results generally consistent with the current understanding of most nova systems, although revealing several interesting systems, there are a number of limitations that are worth of discussion.

\subsection{Inclination}

In systems where the accretion disk has a significant contribution to the luminosity, the magnitude of this contribution is a strong function of the inclination of that disk (or of the system).  \citet{1987MNRAS.227...23W} showed a correlation between the disk inclination and the apparent brightness of nova remnants and the amplitude of their outburst.   The inclination  correction to be applied to the magnitude of a quiescent CN system was initially derived for the dwarf nova U~Geminorum by \citet{1980AcA....30..127P}.  They modeled the disk luminosity in U~Gem by assuming a flat optically thick accretion disk and derived the following relation showing the change in magnitude of the disk as a function of inclination:

\begin{equation}
\Delta M_V(i) = -2.5 \log \left(\cos i+\frac{3}{2}\cos^{2}i\right),\; 0^{\circ}\leq i<90^{\circ}  \textrm{,}
\label{eq:inc}
\end{equation}

\noindent where $i$ is the inclination of the system, defined such that $i=0^{\circ}$ is a polar/face-on system and $i=90^{\circ}$ is edge-on.  This relationship is independent of the disk luminosity and hence the accretion rate and is normalized to the mean disk luminosity (which occurs at $i=56.7^{\circ}$).  We adapt Equation~\ref{eq:inc} such that it may be normalized to any inclination:

\begin{equation}
\Delta M_{v}(i,{\cal I}) = -2.5\log\left( \frac{2\cos{i} + 3\cos^{2} i}{2\cos{{\cal I}} + 3\cos^{2} {\cal I}}  \right),\; 0^{\circ}\leq i,{\cal I}<90^{\circ} \textrm{,}
\label{eq:inc2}
\end{equation}

\noindent where ${\cal I}$ is the inclination at which one chooses as a standard.  The \citet{1980AcA....30..127P} model is applicable for any system in which the accretion disk is the dominant luminosity source, and is therefore only reasonable to apply to MS-Nova systems, the effect of the magnitude change would be expected to decrease significantly for SG-Novae and RG-Novae, i.e. with increasing secondary luminosity.  To investigate the effect of inclination in each nova system we use Equation~\ref{eq:inc2} to minimize the effect of the accretion disk (i.e. edge-on).  However, as the accretion disks in novae are not flat and the secondary has some (non-zero) luminosity in $V$ \citep[original assumptions from][]{1980AcA....30..127P} we adopt an inclination ${\cal I}=80^{\circ}$ to be indicative of the minimum $V$-band flux of an edge-on CN system.  

\begin{figure}
\begin{center}
\includegraphics[width=0.95\columnwidth]{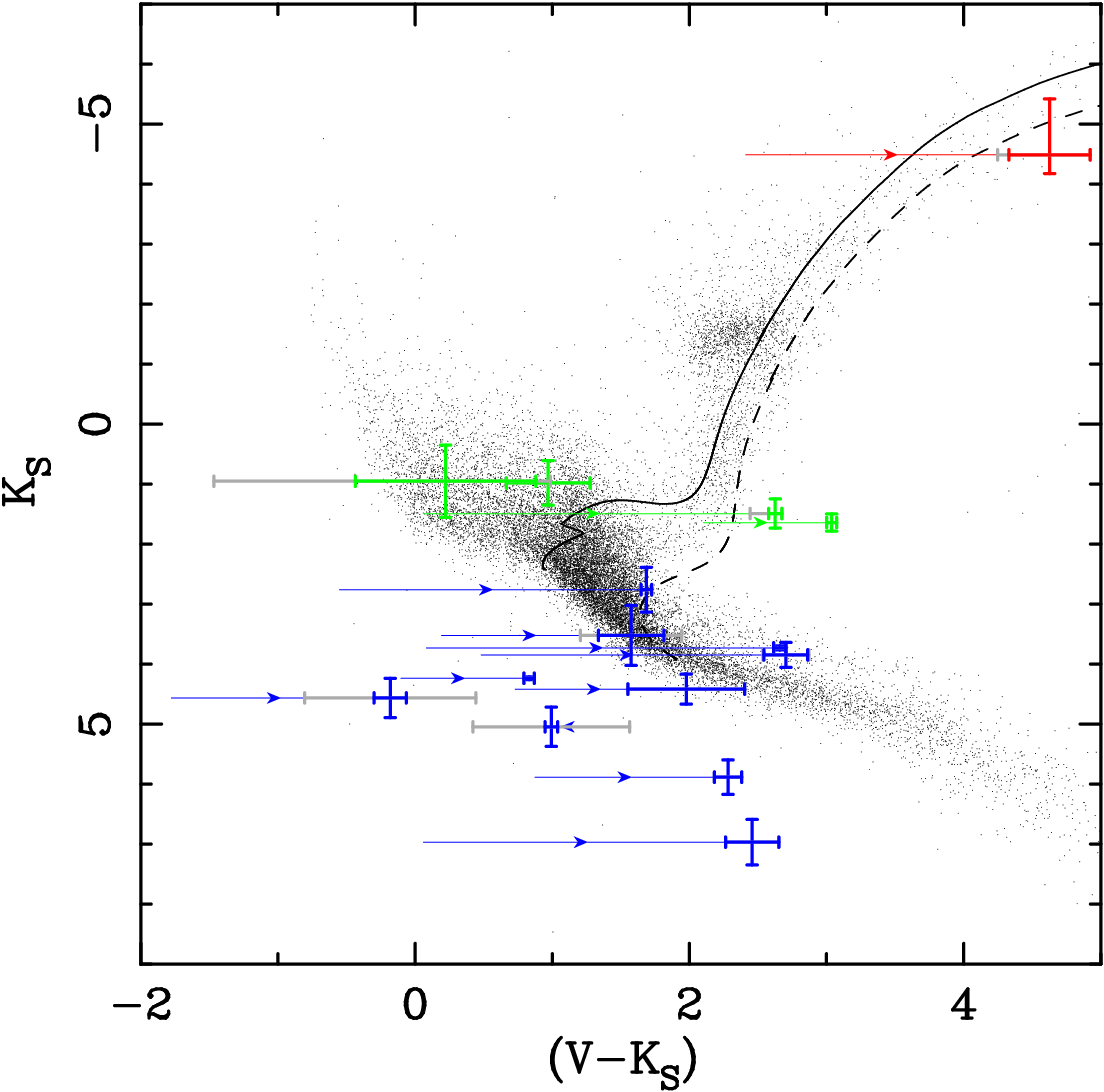}
\end{center}
\caption{As Figure~\ref{fig:one}, but only plotted are quiescent novae for which the system inclination is known.  Each data point shows the accretion disk-corrected position of each system, assumed to be the photometry of only the secondary star.  The gray error bars surrounding some objects show the additional effect of the inclination uncertainty, if known.  The arrowheads indicate the movement of each point from the quiescent photometry position to the accretion disk-corrected position.}
\label{inclination}
\end{figure}

Therefore, to estimate the absolute magnitude $M_V$ of the secondary in a given system at quiescence, it follows that:

\begin{equation}
M_{V}(i,{\cal I}) = V - 5 \log d + 5 - A_V - \Delta M_V(i,{\cal I}) \textrm{,}
\end{equation}

\noindent where $V$ is the apparent magnitude, $d$ distance to the system in (parsecs) and $A_V$ the extinction correction.   However, such an approach is only likely to be valid when it can be assumed that the accretion disk dominates the luminosity and the accretion rate is low (hence the disk is approximately flat).  Hence in SG-Novae and particular RG-Novae systems, where the flux contribution of the secondary is significant, we would expect the effect of system inclination to be substantially reduced.  Additionally, the luminosity ratio of the accretion disk to the secondary will be at its smallest for the $K_{S}$ data.  Hence, if we assume that the $K_{S}$ flux of each quiescent nova system is due only to the secondary, we can produce a color magnitude diagram showing the location of just the secondary star in each system (see Figure~\ref{inclination}).

As can be seen in Figure~\ref{inclination} this very simple approach to inclination has the expected effect of repositioning most MS-nova systems redwards, towards the true position of the secondary star on the main sequence.  However, in order to apply such a technique to all novae, more detailed models of the disks in these systems are required.  Such attempts also suffer from the difficulty in obtaining, and hence lack of, accurate inclination measurements of most nova systems. 

\subsection{Distance}

Another, and probably the most significant complication, comes from the distance determination for each nova system.  As discussed by \citet{2011arXiv1104.3482D} in relation to V2491~Cyg, an incorrect determination of the distance to a nova system can make the difference between suspecting that a system is a MS-nova or a SG-nova.  The distances to a small proportion of novae are well known; i.e. those distances determined by expansion parallax, or those extragalactic novae in galaxies with well determined distances (e.g. SMC, LMC and M31).  However, the distance to a large number of novae has only been determined via the nova ``maximum magnitude versus rate of decline'' (MMRD) relationship.  The MMRD has been shown to contain an inherent 0.6~mag scatter.  However, this has essentially a negligible effect in the analysis undertaken in this paper.  The accuracy of the MMRD is also reliant upon sufficient photometry being taken around the optical maximum of any nova \citep[see e.g.][]{2010ApJ...724..480H}, which becomes more important the faster the outburst is, and adequate monitoring of the decline.  More troublesome though, are the observations that the MMRD performs poorly for the very fastest novae and may not be valid for some recurrent novae \citep{2010ApJS..187..275S} and those with complex lightcurves like V2491~Cyg \citep{2011arXiv1104.3482D}.

Our work has emphasized the lack of reliable distance determinations towards the vast majority of Galactic novae.

\subsection{Extinction}

The effect of extinction towards each nova system upon the work reported in this paper is typically small.  As can be seen in Table~\ref{tb:two}, the extinction towards most nova systems is relatively small and as such, uncertainties in the values have negligible effects on the color-magnitude diagrams presented, especially the NIR plots.  The effect of large extinctions or large uncertainties on extinctions is discussed for the specific example of V1721~Aql by \citet{2011arXiv1104.3068H}.

\subsection{Orbital Period}

\begin{figure}
\begin{center}
\includegraphics[width=0.95\columnwidth]{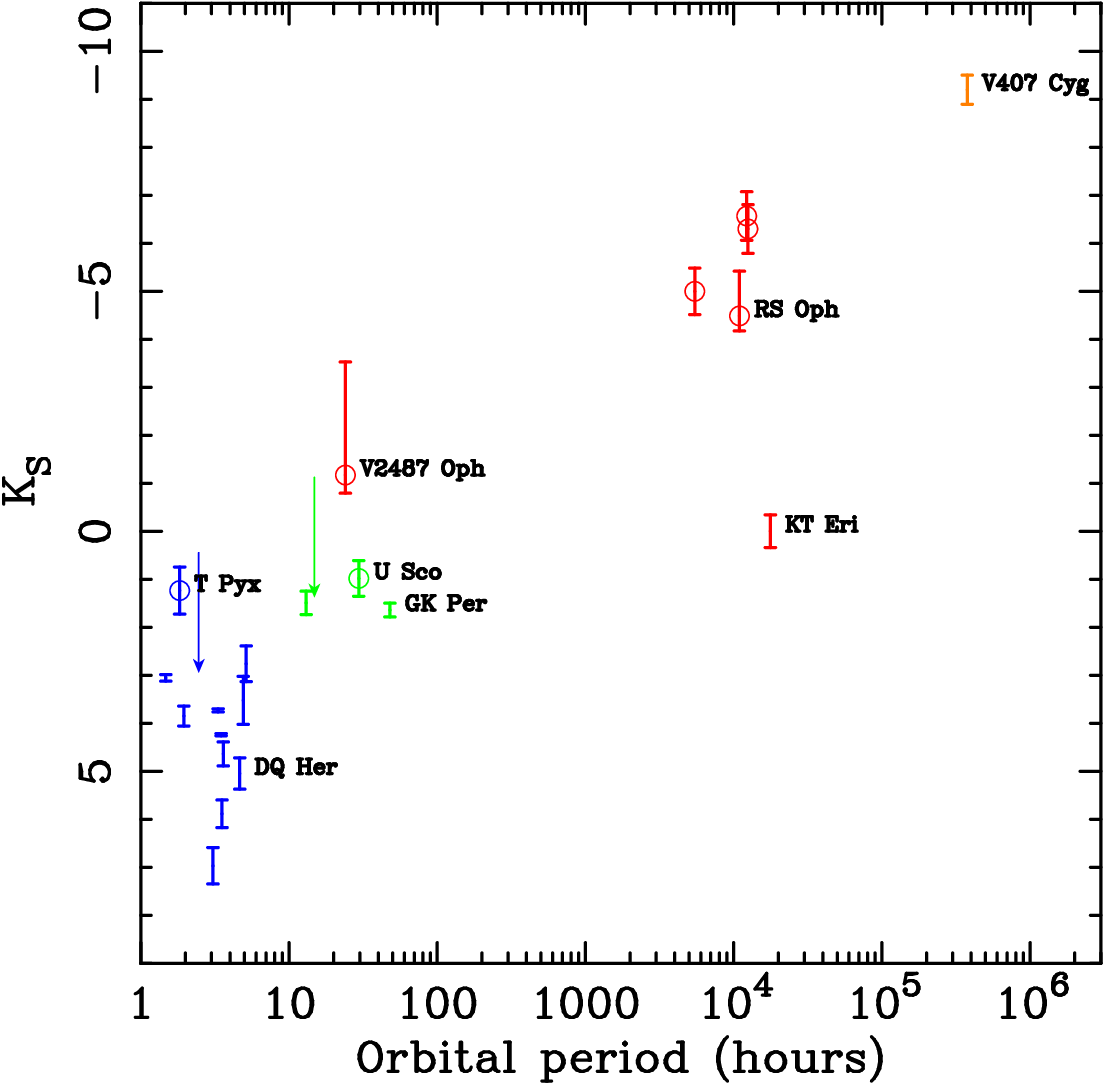}
\end{center}
\caption{Plot showing the known orbital period of quiescent nova systems as a function of their $K_{S}$-band luminosity.  Color-key, as Figure~\ref{fig:one}, a number of note-worthy novae are labeled.  It should be noted that these authors expect the orbital period of V2487~Oph to be much longer (see text for discussion). }
\label{orbitalperiod}
\end{figure}

In Figure~\ref{orbitalperiod} we present a plot of the known orbital periods of quiescent nova systems against that system's quiescent $K_{S}$-band luminosity. The distribution of the data shown in this plot is broadly as expected; the lower the $K_{S}$-band luminosity of the system, the shorter the orbital period.  That is, the systems with the more evolved secondaries must have larger orbital separations between the primary and the secondary.  However, there are a number of interesting outliers to the general trend:

V2487~Oph appears particularly over-luminous for such a short ($\sim 1$ day) period, based on the other RG-Nova systems we would expect a significantly longer period.  It is interesting to note that although \citet{2010ApJS..187..275S} indicates an orbital period of $\sim1$~day for V2487~Oph, he also states ``no significant periodic modulation was seen (in the light curve)''.  As such, we would strongly recommend additional observations in order to directly determine the orbital period of this system.

Although the orbital period of the RG-Nova KT~Eri is similar to others in that group, its luminosity is significantly lower.  As such, we expect the secondary in this system to be less-evolved and hence physically smaller than, for example, RS~Oph.  This secondary must be far from being Roche-lobe filling as was also suggested by \citet{RibKTTEri} based on the period-folded quiescent light curve.

\section{Conclusions}\label{conc}

Our study of the archival optical and NIR photometry of a sample of quiescent Galactic classical and recurrent novae have led us to the following primary conclusions:

\begin{itemize}
\item We have proposed a new classical nova classification system, based solely on the evolutionary state of the secondary star; a main sequence star (MS-Nova), a sub-giant star (SG-Nova) or a red-giant branch star (RG-Nova).
\item The position of a quiescent novae on optical or NIR color-magnitude diagrams is a strong function of both the evolutionary state of the secondary star and the luminosity of any accretion disk.  The more evolved the secondary star, the weaker the position dependency is on the accretion disk.
\item A standard ($V$ versus $(B-V)$) color-magnitude diagram is useful for differentiating between the MS-Novae (which all lie bluer than the main sequence) and the SG-/RG-Novae (which all lie on, or are redder, than the main sequence).  Although this plot can be strongly sensitive to the accretion rate.
\item Additionally NIR color-magnitude diagrams may be used to differentiate between all three sub-classes of secondaries.
\end{itemize}

We also make a number of predictions and observations about systems included in this paper:

\begin{itemize}
\item The recurrent nova CI~Aql - often grouped with T~Pyx - is in fact a SG-Nova, a system more akin to U~Sco.
\item The recurrent nova V2487~Oph is a system similar to KT~Eri, containing a low-luminosity red giant branch star.  This contradicts the short orbital period reported in \citet{2010ApJS..187..275S} and we recommend follow-up observations to directly determine the orbital period.  Additionally, we cannot rule out the possibility that the secondary in this system is a horizontal branch star.
\item Based on the high quiescent luminosity and long orbital period, we predict that the DI~Lac system may contain a sub-giant secondary star.
\item In line with \citet{1994MNRAS.266..761W} and \citet{2003MNRAS.340.1011W}, we propose that EU~Sct contains a luminous red-giant secondary.
\end{itemize}

Follow-up (quiescent) spectroscopic observations of novae, such as those initially reported in \citet{MomayConf}, can greatly aid in the classification of the secondary stars in these systems.

\acknowledgements

This publication makes use of data products from the Two Micron All Sky Survey, which is a joint project of the University of Massachusetts and the Infrared Processing and Analysis Center/California Institute of Technology, funded by the National Aeronautics and Space Administration and the National Science Foundation.  This research has made use of NASA's Astrophysics Data System.  This research has made use of the VizieR catalog access tool, CDS, Strasbourg, France.  VARMR was supported by an STFC PhD studentship and is currently supported by a South African SKA Fellowship.  RAH acknowledges PhD funding from STFC.  The authors would like to thank the referee, Massimo Della Valle, for his very constructive comments.

\clearpage

\begin{landscape}
\begin{table*}
\caption{Optical and NIR photometry of quiescent nova systems\label{tb:one}}
\begin{center}
\begin{tabular}{lllllllllll}
\hline
\hline
Nova          & System            & Recurrent       & Secondary & $B$           & $V$                 & $R$          & $I$                 & $J$              & $H$              & $K_{s}$ \\
Name          & Type (1)          &                 & Spectral  & (mag)         & (mag)               & (mag)        & (mag)               & (mag)            & (mag)            & (mag) \\
              &                   &                 & Class (2) &               &                     &              &                     & \multicolumn{3}{c}{2MASS photometry (3)} \\
\hline
CI Aql        & SG-Nova (4)       & \checkmark      & K-M IV    & 17.10 (5)     & 16.10 (5)           & 15.34 (5,6)  & -                   & $13.67\pm0.04$   & $>13.33\pm0.08$  & $>12.7$ \\
V356 Aql      & MS-Nova           &                 &           & 18.50 (6)     & 18.30 (6)           & 17.80 (6)    & -                   & -                & -                & - \\
V528 Aql      & MS-Nova           &                 &           & 19.25 (6)     & 18.48 (6)           & 17.92 (6)    & -                   & -                & -                & - \\
V603 Aql      & MS-Nova           &                 &           & 11.63 (7)     & 11.67 (7)           & -            & -                   & $11.70\pm0.03$   & $11.51\pm0.03$   & $11.35\pm0.03$ \\
V1229 Aql     & MS-Nova           &                 &           & 19.73 (6)     & 18.59 (6)           & 17.74 (6)    & -                   & $16.3\pm0.1$     & $15.7\pm0.1$     & $>15.5$ \\
V1721 Aql     & SG-Nova (8)       &                 &           & -             & -                   & -            & -                   & $16.6\pm0.2$ (8) & $15.5\pm0.1$ (7) & $14.7\pm0.1$ (8) \\
T Aur         & MS-Nova           &                 &           & 15.20 (7)     & 14.92 (7)           & -            & -                   & $14.01\pm0.03$   & $13.74\pm0.03$   & $13.58\pm0.03$ \\
IV Cep        & MS-Nova           &                 &           & 17.02 (6)     & 16.40 (6)           & -            & -                   & $14.96\pm0.04$   & $14.63\pm0.07$   & $14.30\pm0.08$ \\
V394 CrA      & SG-Nova           & \checkmark      & K         & 19.20 (5)     & 18.40 (5)           & -            & -                   & -                & -                & - \\
T CrB         & RG-Nova           & \checkmark      & M3 III    & 11.60 (5)     & 10.10 (5)           & -            & -                   & $6.00\pm0.02$    & $5.15\pm0.04$    & $4.81\pm0.02$ \\
V407 Cyg      & Sym-Nova          &                 &           & -             & -                   & -            & -                   & $5.70\pm0.02$    & $4.36\pm0.02$    & $3.2\pm0.3$ \\
V476 Cyg      & MS-Nova           &                 &           & 17.24 (7)     & 17.09 (7)           & 17.24 (6,7)  & -                   & $16.0\pm0.1$     & $15.6\pm0.1$     & $15.6\pm0.2$ \\
V1500 Cyg     & MS-Nova           &                 &           & 17.44 (7)     & 17.06 (7)           & 15.15 (6)    & -                   & -                & -                & - \\
V1974 Cyg     & MS-Nova           &                 &           & 16.88 (9)     & 16.63 (9)           & -            & -                   & $15.67\pm0.07$   & $15.3\pm0.1$     & $15.2\pm0.2$ \\
V2491 Cyg     & SG-Nova (10)      &                 &           & 18.34 (11)    & 17.88 (11)          & 17.49 (10)   & 17.14 (11)          & $16.7\pm0.2$ (10) & $16.6\pm0.3$ (10) & $16.7\pm0.6$ (10) \\
HR Del        & MS-Nova           &                 &           & 12.15 (7)     & 12.11 (7)           & -            & -                   & $12.32\pm0.02$   & $12.28\pm0.02$   & $12.22\pm0.03$ \\
KT Eri        & RG-Nova (12)      &                 &           & $15.2\pm0.3$ (12) & -               & $15.3\pm0.3$ (12) & $14.7\pm0.3$ (12) & $14.62\pm0.03$ & $14.16\pm0.05$  & $14.09\pm0.07$ \\
DN Gem        & MS-Nova           &                 &           & 15.76 (7)     & 15.56 (7)           & 15.54 (6,7)  & -                   & $15.43\pm0.05$   & $15.24\pm0.09$   & $15.3\pm0.2$ \\
DQ Her        & MS-Nova (IP)      &                 &           & 14.59 (7)     & 14.45 (7)           & -            & -                   & $13.60\pm0.03$   & $13.28\pm0.04$   & $13.09\pm0.04$ \\
V533 Her      & MS-Nova           &                 &           & 15.78 (7)     & 15.60 (7)           & -            & -                   & $14.71\pm0.03$   & $14.65\pm0.05$   & $14.6\pm0.1$ \\
CP Lac        & MS-Nova           &                 &           & 16.05 (6,7)   & 15.76 (7)           & 15.57 (6,7)  & -                   & $15.03\pm0.06$   & $14.73\pm0.07$   & $14.8\pm0.1$ \\
DK Lac        & MS-Nova           &                 &           & 16.83 (6)     & 17.75 (6)           & 16.59 (6)    & -                   & $16.4\pm0.1$     & $16.0\pm0.2$     & $>15.9$ \\
DI Lac        & SG-Nova (4)       &                 &           & 14.93 (6)     & 14.68 (6)           & -            & -                   & $13.73\pm0.04$   & $13.54\pm0.05$   & $13.40\pm0.04$ \\
IM Nor        & MS-Nova           & \checkmark      &           & 19.30 (5)     & 18.30 (5)           & -            & -                   & $15.46\pm0.09$   & $>13.6$          & $>13.4$ \\
RS Oph        & RG-Nova           & \checkmark      & M0/2 III  & $12.30\pm0.03$ (13,14) & $10.98\pm0.03$ (13,14) & $9.50\pm0.03$ (13,14) & $8.67\pm0.02$ (13,14) & $7.64\pm0.02$    & $6.86\pm0.04$    & $6.50\pm0.02$ \\
V849 Oph      & MS-Nova           &                 &           & 19.30 (6)     & 18.80 (6)           & 18.50 (6)    & -                   & -                & -                & - \\
V2487 Oph     & RG-Nova (4)       & \checkmark      &           & 18.10 (5)     & 17.30 (5)           & -            & -                   & $15.36\pm0.08$   & $14.9\pm0.1$     & $14.41\pm0.09$ \\
GK Per        & SG-Nova (IP)      &                 &           & 13.81 (7)     & 13.02 (7)           & -            & -                   & $10.86\pm0.02$   & $10.27\pm0.02$   & $10.06\pm0.02$ \\
RR Pic        & MS-Nova           &                 &           & 12.18 (7)     & 12.21 (7)           & -            & -                   & $12.46\pm0.02$   & $12.40\pm0.02$   & $12.25\pm0.02$ \\
CP Pup        & MS-Nova           &                 &           & 15.17 (7)     & 14.97 (7)           & -            & -                   & $14.34\pm0.03$   & $14.24\pm0.04$   & $14.02\pm0.07$ \\
T Pyx         & MS-Nova           & \checkmark      &           & 15.60 (5)     & 15.60 (5)           & -            & -                   & $15.15\pm0.05$   & $14.96\pm0.09$   & $14.67\pm0.09$ \\
U Sco         & SG-Nova           & \checkmark      & K2 IV     & 18.55 (15)    & 18.01 (15)          & 17.67 (15)   & 17.21 (15)          & $16.8\pm0.1$     & $16.4\pm0.2$     & $>15.2$ \\
V745 Sco      & RG-Nova           & \checkmark      & M6 III    & 20.50 (5)     & 18.70 (5)           & -            & -                   & $10.04\pm0.03$   & $8.85\pm0.04$    & $8.26\pm0.04$ \\
EU Sct        & RG-Nova (4)       &                 &           & 20.02 (6)     & 17.37 (6)           & 18.80 (6)    & -                   & $12.26\pm0.05$   & $11.18\pm0.04$   & $10.80\pm0.04$ \\
FH Ser        & MS-Nova           &                 &           & 16.18 (7)     & 15.06 (7)           & 14.79 (6,7)  & -                   & $15.25\pm0.06$   & $15.0\pm0.1$     & $14.7\pm0.1$ \\
V3890 Sgr     & RG-Nova           & \checkmark      & M5 III    & 16.40 (5)     & 15.50 (5)           & -            & -                   & $9.83\pm0.02$    & $8.77\pm0.03$    & $8.26\pm0.02$ \\
LV Vul        & MS-Nova           &                 &           & 16.56 (6)     & 15.84 (6)           & 15.24 (6)    & -                   & -                & -                & - \\
NQ Vul        & MS-Nova           &                 &           & 18.27 (6)     & 17.22 (6)           & 16.32 (6)    & -                   & $15.03\pm0.05$   & $14.63\pm0.06$   & $14.26\pm0.08$ \\
\hline
LMCN 2009-02a & SG-Nova (16)      & \checkmark (16) &           & $18.3\pm0.4$(16) & -                & $19.2\pm0.6$ (16) & -              & -                & -                & -\\
M31N 2007-12b & RG-Nova (13)      &                 &           & -             & 24.61$\pm$0.09 (13) & -            & 22.33$\pm$0.04 (13) & -                & -                & -\\
\hline
\end{tabular}
\end{center}
1.~Standard classification for each system, unless otherwise stated, 2.~Spectral classes of the secondary stars of known RNe from \citet{2008ASPC..401...31A}, 3.~Unless otherwise stated, all NIR photometry is taken directly from the 2MASS Catalog \citep{2006AJ....131.1163S}, 4.~This paper, 5.~\citet{2010ApJS..187..275S}, 6.~\citet{1994AJ....108..639S}, 7.~\citet{1994A&AS..104...79B}, 8.~\citet{2011arXiv1104.3068H}, 9.~\citet{2002AIPC..637..323S}, 10.~\citet{2011arXiv1104.3482D}, 11.~\citet{2010arXiv1009.0822M}, 12.~\citet{RibKTTEri}, 13.~\citet{2009ApJ...705.1056B}, 14.~\citet{2008ASPC..401..203D}, 15.~\citet{2010AJ....140..925S}, 16.~Bode et al. in prep.
\end{table*}
\clearpage
\end{landscape}

\end{document}